\documentclass[aps,prl,reprint,showpacs,floatfix]{revtex4-2}
\usepackage{amsmath,amssymb,graphicx}

\begin{document}

\title{Finite-Inertia Corrections and Breakdown of Gor'kov Theory in Acoustic Levitation of Droplets}

\author{Hollis Williams}
\affiliation{Theoretical Sciences Visiting Program, Okinawa Institute of Science and Technology Graduate University, Onna-son, Okinawa 904-0495, Japan}

\begin{abstract}
Acoustic levitation is widely used for contactless droplet manipulation, yet the
standard Gor'kov description obtained by time averaging the acoustic
field lacks a quantitative validity criterion.  In this work, we derive
Gor'kov theory as the leading-order slow time limit of the instantaneous
radiation force, compute the first finite-inertia correction, and obtain a
simple breakdown parameter.
The correction reduces the effective trapping drift and predicts fast time
oscillations of amplitude $x_1^{\mathrm{max}}\sim\lambda/8$, corresponding to
hundreds of micron for typical ultrasonic levitation experiments. This sets a
measurable criterion for experiments using phased transducer arrays. Our
results provide a universal rule of thumb for acoustic trap design and clarify
where time-averaged radiation force models fail.

\end{abstract}

\maketitle

\section{Introduction}
Contactless manipulation of liquid droplets and particles using acoustic fields has emerged as a powerful tool across microfluidics, materials processing, and soft-matter physics, enabling trapping, transport, and assembly without physical contact or contamination \cite{marx, ozcelik, mohanty, dai}.  Recent advances in phased transducer arrays and programmable acoustic fields have further expanded the scope of these techniques, allowing dynamic reconfiguration of traps and complex droplet trajectories in air and liquids \cite{marzo, cox, agyri, contreras,  sun,   zehn,   meng}.  The theoretical description of acoustic trapping is, however, still largely based on time-averaged radiation forces. Since the seminal work of Gor’kov, it has been standard to replace the rapidly oscillating acoustic field by an effective potential obtained by averaging over the acoustic period \cite{gorkov, settnes, wang}.  This approach has proven extremely successful and underpins most analytical modeling and numerical design tools currently used in acoustic levitation and manipulation \cite{mitri,  dual,simon,  yang, castro}.

Despite this success, the validity of time-averaged descriptions is rarely examined explicitly. In practice, experimental implementations often rely on empirical tuning of driving parameters, trap geometries, or control protocols when stable trapping or transport fails, particularly for liquid droplets with finite inertia \cite{watanabe, hasegawa, brown}.  In parallel, numerical simulations typically invoke time averaging from the outset in order to make the problem tractable, precluding a systematic assessment of when such approximations may break down \cite{chen, xie}.  In this Letter, we address this gap by developing the simplest dynamical model of acoustic droplet levitation in which time averaging is not assumed a priori. We consider the instantaneous radiation force acting on a droplet with finite inertia in a standing acoustic field and analyze its dynamics using a systematic multiple-scale expansion. Within this framework, the familiar Gor’kov radiation force emerges naturally as the leading-order result in the limit of separation of time scales, rather than as an imposed assumption.

Going beyond this limit, we derive explicit corrections that quantify when and where the time-averaged description breaks down. We show that the onset of breakdown is intrinsically position dependent, occurring first in regions of large force gradients rather than near the extrema of the time-averaged potential. Moreover, the magnitude of the fast-time oscillations is set by a dimensionless parameter and constitutes a fixed fraction of the trap size, implying that these effects cannot be eliminated by tuning material properties alone.

These results have immediate implications for contemporary acoustic manipulation platforms, including phased transducer arrays used for droplet handling, transport, and assembly \cite{seah}.  Even in regimes where time-averaged calculations predict stable trapping, our analysis identifies spatial regions in which fast-time oscillations are expected to induce jitter, drift, or loss of control \cite{murata}. More broadly, the present work clarifies the limits of applicability of time-averaged radiation force models and provides a universal framework for assessing their validity across a wide range of acoustic manipulation systems.

\section{Acoustic Trapping Dynamics}

We consider the simplest dynamical setting in which the standard time-averaging assumption underlying acoustic radiation forces can be examined.  More specifically, we study the center-of-mass motion of a liquid droplet levitated in an acoustic standing wave and neglect internal shape deformations. As discussed below, this reduction is controlled by a separation of timescales between translational motion and internal shape modes.

\subsection{Acoustic field}

We begin with a one-dimensional standing acoustic wave of angular frequency $\omega$ and wavenumber $k$, with pressure field
\begin{equation}
p(x,t) = p_0 \cos(kx)\cos(\omega t),
\end{equation}
where $p_0$ is the pressure amplitude. This field represents the local structure of an acoustic trap along its principal direction of confinement.

\subsection{Instantaneous radiation force}

In the long-wavelength limit $kR \ll 1$, the acoustic radiation force acting on a small droplet with radius $R$ arises from gradients in the local acoustic energy density. For a standing wave, the instantaneous force can be written in the generic form
\begin{equation}
F(x,t) = A\,\sin\!\big(2k x\big)\cos^2(\omega t),
\label{eq:force}
\end{equation}
where $A$ sets the overall magnitude of the force and depends on the droplet volume, acoustic amplitude, and material contrast parameters. Importantly, Eq.~\eqref{eq:force} retains the explicit time dependence of the acoustic field and no time averaging has been performed.

\subsection{Equation of motion}

Treating the droplet as a point mass with finite inertia, its center-of-mass motion obeys
\begin{equation}
m\ddot{x}(t) + \gamma \dot{x}(t)
= A\,\sin\!\big(2k x(t)\big)\cos^2(\omega t),
\label{eq:eom_dimensional}
\end{equation}
where $m$ is the droplet mass and $\gamma$ is an effective drag coefficient accounting for viscous dissipation in the surrounding fluid. Equation~\eqref{eq:eom_dimensional} constitutes the minimal dynamical model of the levitated droplet prior to any averaging approximation.  The system under consideration is illustrated in Fig. 1. A liquid droplet is levitated in a one-dimensional standing acoustic field, with the droplet’s mean position denoted $X_0$ and fast oscillations indicated by $\epsilon X_1$.  The separation between slow and fast dynamics forms the basis of the multiple-scale analysis in the next section.

\subsection{Non-dimensionalization}

We next introduce dimensionless variables
\begin{equation}
X = kx, \qquad \tau = \omega t,
\end{equation}
Substituting these in, Eq.~\eqref{eq:eom_dimensional} becomes
\begin{equation}
\epsilon^2 \frac{d^2 X}{d\tau^2}
+ \delta \frac{dX}{d\tau}
= \sin(2X)\cos^2(\tau),
\label{eq:eom_dimensionless}
\end{equation}
where the dimensionless parameters are defined as
\begin{equation}
\epsilon^2 = \frac{m\omega^2}{Ak}, \qquad
\delta = \frac{\gamma\omega}{Ak}.
\end{equation}

The parameter $\epsilon$ compares the acoustic oscillation frequency $\omega$ to the characteristic trapping frequency $\omega_{\mathrm{trap}} \sim \sqrt{Ak/m}$ obtained by linearizing and time-averaging the force near a stable equilibrium. In standard treatments, the assumption $\epsilon \ll 1$ is implicit. Here, by contrast, Eq. 5 retains the full time dependence of the forcing, and no averaging has been assumed at any stage.  The one-dimensional standing-wave field considered here should be understood as the local normal form of the acoustic field near a stable levitation point.  More complex trap geometries modify the spatial dependence of the force but do not alter its temporal structure or the associated separation of timescales.

\begin{figure}

\includegraphics[width=80mm]{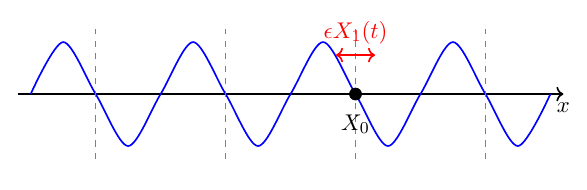}% 

\caption{\label{fig:epsart}Schematic of the theoretical setup. A liquid droplet (denoted by a black dot) levitates in a one-dimensional standing acoustic wave (shown in blue). Its motion is decomposed into a slow mean position $X_0$ and $\epsilon X_1$ driven by the instantaneous acoustic radiation force (shown with red arrows). The analysis does not assume time averaging of the force.  }
\end{figure}

\section{Multiple-scale analysis}

We now analyze the dimensionless equation of motion (5), where $\tau=\omega t$ is the fast acoustic time and $\epsilon\ll 1$ characterizes the separation between the acoustic oscillation timescale and the characteristic timescale for translational motion. 

%\begin{equation}
%\epsilon^2 \frac{d^2 X}{d\tau^2}
%+ \delta \frac{dX}{d\tau}
%= \sin(2X)\cos^2(\tau),
%\label{eq:eom_dimensionless}
%\end{equation}

\subsection{Separation of fast and slow times}

Since a net translational motion arises only through the combined effect of many acoustic cycles, we introduce a slow time
\begin{equation}
T = \epsilon^2 \tau,
\end{equation}
and treat the droplet position as a function of both variables
\begin{equation}
X(\tau) = X(\tau,T).
\end{equation}
Derivatives with respect to $\tau$ transform as
\begin{align}
\frac{d}{d\tau} &= \partial_\tau + \epsilon^2 \partial_T, \\
\frac{d^2}{d\tau^2} &= \partial_\tau^2
+ 2\epsilon^2 \partial_\tau\partial_T
+ \mathcal{O}(\epsilon^4).
\end{align}

\noindent
We seek an asymptotic expansion of the form
\begin{equation}
X(\tau,T) = X_0(T) + \epsilon^2 X_1(\tau,T)
+ \mathcal{O}(\epsilon^4),
\label{eq:expansion}
\end{equation}
where $X_0(T)$ represents the slowly varying mean position of the droplet.

\subsection{Leading-order dynamics}

Substituting Eq.~\eqref{eq:expansion} into Eq.~\eqref{eq:eom_dimensionless} and collecting terms at order $\mathcal{O}(1)$ yields
\begin{equation}
\partial_\tau^2 X_0 = 0.
\end{equation}
It follows that at leading order, the droplet position is independent of the fast time
\begin{equation}
X_0 = X_0(T),
\end{equation}
indicating that no net motion occurs on the acoustic timescale.

\subsection{Higher order}

At order $\mathcal{O}(\epsilon^2)$, we obtain
\begin{equation}
\partial_\tau^2 X_1
= \sin\!\big(2X_0\big)\cos^2(\tau)
- \delta \partial_T X_0.
\label{eq:order_eps2}
\end{equation}

\noindent
Equation~\eqref{eq:order_eps2} admits a bounded, periodic solution for $X_1$ only if the right-hand side has zero mean over the fast timescale. This condition enforces
\begin{equation}
\delta \partial_T X_0
= \left\langle
\sin\!\big(2X_0\big)\cos^2(\tau)
\right\rangle_{\tau},
\label{eq:solvability}
\end{equation}
where $\langle \cdot \rangle_{\tau}$ denotes the average over one period of the fast time $\tau$.  Evaluating the average explicitly
\begin{equation}
\left\langle \cos^2(\tau) \right\rangle_{\tau}
= \frac{1}{2},
\end{equation}
we obtain the effective slow time evolution equation
\begin{equation}
\partial_T X_0
= \frac{1}{2\delta}\sin\!\big(2X_0\big).
\label{eq:slow_dynamics}
\end{equation}

\subsection{Gor'kov force}
Equation~\eqref{eq:slow_dynamics} governs the evolution of the slowly varying droplet position $X_0(T)$ and describes overdamped motion under an effective restoring force. This equation can be written in gradient form as
\begin{equation}
\partial_T X_0 = -\frac{1}{\delta}\,\frac{dU_{\mathrm{eff}}}{dX_0},
\end{equation}
with an effective potential
\begin{equation}
U_{\mathrm{eff}}(X_0) = -\frac{1}{4}\cos(2X_0).
\end{equation}

Restoring dimensions and reverting to the physical coordinate $x$, the corresponding effective force acting on the droplet is
\begin{equation}
F_{\mathrm{eff}}(x)
= -\frac{dU_{\mathrm{eff}}}{dx}
= \frac{A}{2}\sin(2kx),
\end{equation}
which coincides with the standard Gor'kov radiation force obtained by time averaging the acoustic energy density of a standing wave.  This effective force has not been assumed here a priori, but instead emerges as a solvability condition from the multiple-scale expansion of the fully time-dependent equation of motion in the limit $\epsilon\ll1$.

\subsection{Correction to Gor'kov}

Including the first correction beyond leading order, the slow time equation becomes
%\begin{equation}
%\delta \frac{d X_0}{d T} 
%= \frac{1}{2} \sin(2X_0) - \frac{\epsilon^2}{16} %\sin^3(2X_0),
%\end{equation}
%or equivalently
\begin{equation}
\frac{d X_0}{d T} = \frac{1}{2\delta} \sin(2X_0) \left[ 1 - \frac{\epsilon^2}{8} \sin^2(2X_0) \right].
\end{equation}

\noindent
The correction term arises from the quadratic contribution of the fast-time oscillatory component
\begin{equation}
X_1(\tau,T) = - \frac{1}{4} \sin(2X_0) \cos(2\tau)
\end{equation}
and represents a reduction in the effective drift due to finite inertia. It vanishes at the nodes of the acoustic standing wave (\(X_0 = n\pi/2\)) and is maximal near the trap midpoints. This term naturally sets the scale for the breakdown of Gor'kov theory when the slow motion is no longer negligible compared to the acoustic oscillations.  The amplitude of the fast time oscillations varies with the droplet’s slow position $X_0$, as shown in Fig. 2.  $X_1$ is maximized near the midpoints between nodes and anti-nodes, highlighting the regions where the time-averaged approximation is least valid.

\begin{figure}

\includegraphics[width=80mm]{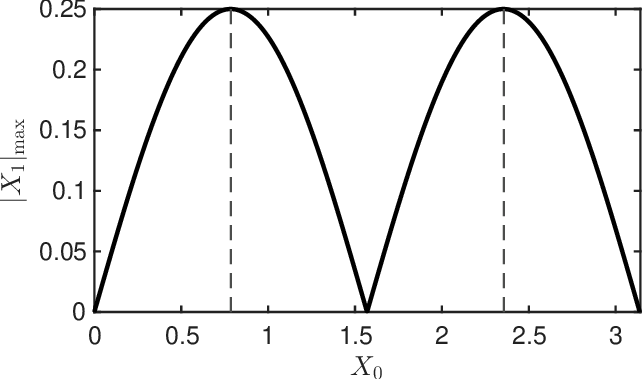}% 

\caption{\label{fig:epsart} Maximum amplitude of the fast-time oscillatory motion, $|X_1|_{\max} = \frac{1}{4}|\sin(2X_0)|$, as a function of the slow time droplet position within a standing wave trap. Fast oscillations are largest near trap midpoints and vanish at nodes, indicating where time-averaged descriptions first break down. }
\end{figure}

\begin{figure}

\includegraphics[width=80mm]{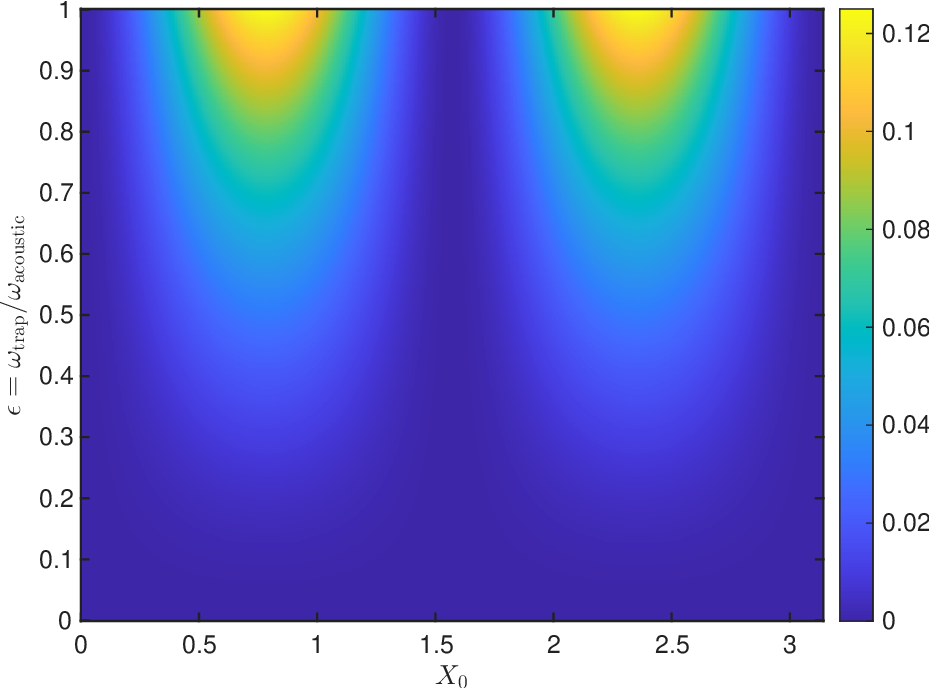}% 

\caption{\label{fig:epsart} Regime diagram for the validity of the time-averaged theory. The color scale shows the dimensionless breakdown parameter $\Lambda (X_0, \epsilon) = \epsilon^2 \sin^2 ( 2 X_0)/8$, which controls the magnitude of finite-inertia corrections to Gor’kov theory. Large values indicate regions where fast-time oscillations become significant and instantaneous dynamics cannot be neglected. The breakdown is intrinsically position dependent and occurs first near trap midpoints. }
\end{figure}

\subsection{Criterion for breakdown of Gor'kov theory}

%The next-order correction to the slow-time equation, 
%\begin{equation}
%\frac{d X_0}{d T} = \frac{1}{2\delta} \sin(2X_0) 
%\left[ 1 - \frac{\epsilon^2}{8} \sin^2(2X_0) \right],
%\end{equation}
Eq. (21) immediately suggests a natural criterion for the validity of Gor'kov theory.  The correction becomes significant when
\begin{equation}
\frac{\epsilon^2}{8} \sin^2(2X_0) \sim 1,
\end{equation}
which corresponds to
\begin{equation}
\epsilon_{\mathrm{crit}} \sim \mathcal{O}(1),
\quad \text{i.e.}\quad
\omega_{\mathrm{trap}} \sim \omega_{\mathrm{acoustic}}.
\end{equation}

\noindent
Physically, this criterion indicates that Gor'kov theory is only valid when the droplet moves a negligible amount during a single acoustic cycle. Breakdown first occurs near the trap midpoints (\(X_0 \sim (n+\tfrac{1}{4})\pi\)) where \(|\sin(2X_0)|\) is maximal, whilst nodes of the standing wave (\(X_0 = n \pi/2\)) remain well described by the leading-order theory. 

Fig 3 shows a contour map of the dimensionless breakdown parameter.   Regions of high color intensity correspond to locations where the instantaneous oscillation amplitude is a large fraction of the trap size, indicating a breakdown of the time-averaged approximation.  It can be seen that breakdown first occurs near the midpoints between nodes and antinodes of the standing wave and not at the extrema of the time-averaged potential, and that the size of these regions grows with increasing $\epsilon$.  

%It follows immediately that in experiments (for example, using transducer arrays) placing droplets in high-gradient regions of the field can lead to significant fast-time excursions or loss of control, even when the time-averaged potential predicts stable trapping.  The universal nature of the dimensionless parameter means that this guidance holds for a wide range of droplet sizes, acoustic amplitudes, and trap frequencies.

\section{Discussion}
Our analysis provides a systematic derivation of Gor'kov theory, its first finite-inertia correction, and a quantitative criterion for breakdown. These results give engineers a predictive rule for trap design and droplet manipulation in transducer arrays, allowing them to anticipate deviations from time-averaged theory. Extensions to 2D/3D arrays and deformable droplets are straightforward \cite{labchip}.  For typical ultrasonic levitation experiments with driving frequencies
$f \sim 20-100\,\mathrm{kHz}$ and acoustic wavelengths
$\lambda \sim 1-10\,\mathrm{mm}$, the predicted fast time oscillation
amplitude $x_1^{\mathrm{max}} \simeq \lambda/8$ corresponds to motion on the
order of $100-1000\,\mu\mathrm{m}$. These values are experimentally resolvable and are comparable to the spatial extent of many acoustic
traps. The breakdown parameter shown in Fig. 3 therefore reaches order unity
under standard operating conditions, consistent with the sensitivity and
irreproducibility problems often encountered by experimentalists working with transducer arrays.

%More broadly, the present results clarify the limits of applicability of time-averaged radiation-force models and provide a systematic framework for predicting when such descriptions are expected to fail.

Our analysis reveals several physically nontrivial features of acoustic trapping dynamics which are not captured by the standard time-averaged description. Firstly, the breakdown of time averaging is intrinsically position dependent: fast time oscillations become largest near the midpoints between nodes and antinodes, and not close to the extrema of the Gor’kov potential. This behavior is not apparent from the conventional picture, which predicts the strongest restoring forces close to trapping nodes.  Secondly, the onset of breakdown is directly linked to spatial gradients of the acoustic field \cite{chu, doinikov}.  The amplitude of fast oscillations is controlled by the local curvature of the instantaneous radiation force, rather than its time-averaged value. This establishes a direct connection between the validity of Gor’kov theory and experimental features such as trap geometry and field gradients.

Finally, the magnitude of the fast oscillations is set by a dimensionless ratio and constitutes a fixed fraction of the trap size. As a result, these corrections cannot be eliminated by tuning droplet properties or acoustic contrast factors.  They can only be removed by changing the separation of time scales (for example, through changes in driving frequency). This universality underscores that the breakdown mechanism we have found is generic and not system-specific \cite{nabavi}.

%These results provide a natural explanation for the sensitivity and irreproducibility often reported in acoustic levitation experiments, particularly those employing phased transducer arrays. Even when time-averaged models predict stable trapping, finite-inertia effects can induce large fast-time oscillations in specific spatial regions, leading to jitter, drift, or sudden loss of levitation. The strong position dependence of the breakdown criterion implies that nominally equivalent trap locations can exhibit markedly different stability, consistent with the empirical tuning and trial-and-error commonly employed in experimental implementations.

\begin{acknowledgments}

\noindent
This research was conducted whilst the author was visiting the Okinawa Institute of Science and Technology (OIST) through the Theoretical Sciences Visiting Program (TSVP). 
\end{acknowledgments}

\end{document}